\documentclass{kapproc} 
\usepackage{psfig,epsf}
\normallatexbib

\begin{document}

\articletitle{Nishimori point in random-bond Ising and Potts models in
2D}

\author{A.~Honecker}
\affil{Institut f\"ur Theoretische Physik, TU Braunschweig\\
    Mendelssohnstr.\ 3, 38106 Braunschweig, Germany}
\email{a.honecker@tu-bs.de}
\author{J.~L.~Jacobsen}
\affil{LPTMS, Universit\'e Paris-Sud \\
       B\^atiment 100, 91405 Orsay, France}
\email{jacobsen@ipno.in2p3.fr}
\author{M.~Picco}
\affil{LPTHE, Universit\'es Paris VI et Paris VII\\
        4 place Jussieu, 75252 Paris C\'edex 05, France}
\email{picco@lpthe.jussieu.fr}
\author{P.~Pujol}
\affil{Laboratoire de Physique, Groupe de Physique Th\'eorique\\
ENS Lyon, 46 All\'ee d'Italie, 69364 Lyon C\'edex 07, France}
\email{Pierre.pujol@ens-lyon.fr}

\begin{abstract}
We study the universality class of the fixed points of the 2D random
bond $q$-state Potts model by means of numerical transfer matrix
methods. In particular, we determine the critical exponents associated
with the fixed point on the Nishimori line. Precise measurements
show that the universality class of this fixed point is inconsistent
with percolation on Potts clusters for $q=2$, corresponding to the Ising
model, and $q=3$.
\end{abstract}

\begin{keywords}
Spin glass, Potts model, Nishimori line, conformal field theory
\end{keywords}

\section*{Introduction}

During the last decade, the study of disordered systems has attracted
much interest. This is true in particular in two dimensions, where the
possible types of critical behavior for the corresponding pure models
can be classified using conformal field theory \cite{BPZ}. Recently,
similar classification issues for disordered models have been
addressed through the study of random matrix ensembles \cite{Z}, but
many fundamental questions remain open.

An important category of 2D disordered systems is given by models
where the disorder couples to the local energy density (random Potts
models).  Here we shall study such models that interpolate between
ferromagnetic random bond disorder, and a stronger $\pm J$ type
disorder.  Our main focus shall be on the cases with $q=2$ (Ising) or
$q=3$ states.

\section{Phase Diagram}
The Ising model on a square lattice is one of the
most popular two-dimensional systems. It is specified by the energy of
a spin configuration
\begin{equation}
{\cal H}(\{S_i\}) = \sum_{\langle i,j \rangle} J_{ij} \delta_{S_i,S_j}  \, ,
\label{RBI}
\end{equation}
where the sum is over all bonds and the coupling constants $J_{ij}$
are bond dependent. Different distributions of disorder can be
considered. The most common ones are the $J_{ij}=\pm 1$ and the
Gaussian distribution of disorder. In this work we will study in
particular the $J_{ij}=\pm 1$ Random-Bond Ising Model (RBIM) with the
following probability distribution:
\begin{equation}
P(J_{ij}) = p \delta(J_{ij}-1) + (1-p) \delta(J_{ij}+1) \, .
\label{pmJprob}
\end{equation}
The topology of the phase diagram of the RBIM depends crucially on the
type of disorder one considers. An instructive example is provided by
a disorder having only two possible values for the bonds with equal
signs and probabilities. It is by now well established \cite{JC} that
the only non-trivial fixed points are located at the extrema of the
boundary of the ferromagnetic phase, corresponding to the pure Ising
fixed point and a zero temperature fixed point which turns out to be
in the percolation universality class.

\begin{figure}[t]
\centerline{\psfig{file=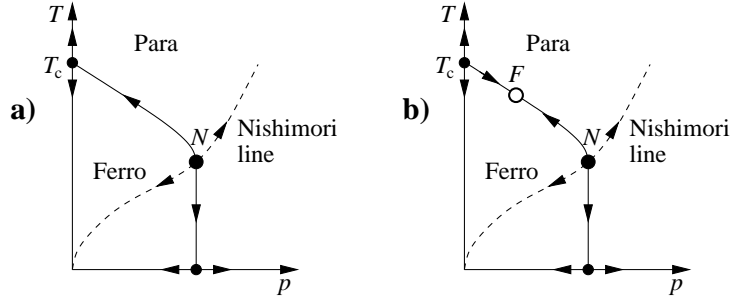,width=0.8\columnwidth}}
\caption{
Phase diagram of the two-dimensional $\pm J$ random-bond Ising (a)
and $q > 2$ Potts (b) models. Note the additional ferromagnetic fixed
point {\em F} in the Potts model (b).
\label{phaseRBI}
}
\end{figure}

When the distribution contains also bonds with different signs (like
in (\ref{pmJprob})), the situation is more subtle. In particular, it
is known since the work of McMillan \cite{MM} that there exists an
unstable fixed point at finite temperature and a finite value of
disorder $p_c$ and another fixed point at zero temperature and a value
of disorder $p \simeq p_c$ (McMillan obtained these results with a
Gaussian distribution of disorder). Thus for the RBIM, one expects
three fixed points (see Fig.~1a): i) the fixed point corresponding to
the case without disorder. Close to this point, one expects that the
physics is just described by the usual perturbation of the Ising model
with weak disorder \cite{DotDot,shalaev,shankar,ludwig2}, {\it i.e.}\
one flows back to the model without disorder. ii) A random fixed point
$N$ at finite temperature and a finite value of disorder. Describing
this unstable fixed point is the main purpose of this work. Since this
fixed point is unstable under two parameters (temperature and
disorder) it is very difficult to study numerically. We will come back
to this point later. iii) Finally, there is a third fixed point at
zero temperature but non-vanishing disorder. The universality class of
this last fixed point is also unknown at present.

For the more general case of the random $q$-state Potts model (RBPM)
with $q>2$, the situation is slightly more complicated. We define
this model by
\begin{equation}
 {\cal H}(\{S_i\})
     = -\sum_{\langle i,j \rangle} \delta^{(q)}(S_i-S_j+J_{ij})  \, ,
 \label{PGG}
\end{equation}
where $S_i=1,2,\ldots,q$, and $\delta^{(q)}(x)=1$ if $x=0 \mbox{ mod }
q$ and zero otherwise. The randomness now takes the form of a local
``twist'' $J_{ij}$, which is clearly a more severe type of disorder
than simple bond randomness. The variables $J_{ij}$ are taken from the
distribution
\begin{equation}
 P(J_{ij}) = \big(1-(q-1)p \big) \delta(J_{ij}) +
              p \sum_{J=1}^{q-1} \delta(J_{ij}-J) \, ,
\label{olddist}
\end{equation}
with $0 \le p \le 1/(q-1)$ controlling the strength of the
randomness. We shall refer to this model, which was originally
introduced in Ref.~\cite{ns}, as the Potts Gauge Glass (PGG). The
particular form of the randomness ensures the existence of a Nishimori
line (see below).  For $q=2$, the PGG reduces to the RBIM, and for
$p=1/q$ it was studied analytically in \cite{Gold}. The PGG is also
connected to the RBPM: When $q>2$ the pure Potts model ($p=0$) should
be {\em unstable} to a small amount of randomness, meaning that the
renormalization group (RG) flow cannot be as indicated in
Fig.~\ref{phaseRBI}a. Indeed, the existence of a new fixed point {\it
F} between the pure and random fixed points (see open circle in
Fig.~\ref{phaseRBI}b) is now well established analytically
\cite{ludwig2,Perturb} and numerically for $q=3$ \cite{Sorensen}. If $(q-2)$,
and hence the value of $p$ at {\it F}, is sufficiently small,
frustration effects are negligible, and we should flow to the {\em
same} random fixed point for any kind of randomness. For reasons of
continuity we expect this argument to hold true also for higher values
of $q$ \cite{JC}.

As mentioned above, a numerical study of the random fixed point $N$ at
finite temperature and finite disorder is a difficult task since it is
unstable in both of its parameters. Fortunately, for a certain class
of probability distributions, Nishimori has shown that a so-called
`Nishimori' line exists where many properties can be calculated
exactly \cite{N}. For the RBIM with the probability distribution
(\ref{pmJprob}), this line is given by
\begin{equation}
{\rm e}^{\beta} =  {1-p \over p} \, ,
\label{NishimoriLine}
\end{equation}
with $\beta = 1/T$. On the Nishimori line, the internal energy can be
calculated exactly and an upper bound can be given for the specific
heat. Also of interest is an equality of the moments of the spin
correlation functions (see below). Nishimori has further proved
inequalities for the correlation functions which yield important
constraints on the topology of the phase diagram which is shown in
Fig.~\ref{phaseRBI}a for the $\pm J$ RBIM. Since the Nishimori line is
also invariant under Renormalization Group (RG) transformations
\cite{LG}, the intersection of the Nishimori line and the Ferro-Para
transition line must be a fixed point which is identified with the
random fixed point $N$. This so-called Nishimori point corresponds to
a new universality class. The full line in the phase diagram
Fig.~\ref{phaseRBI}a is the phase boundary between the ferromagnetic
and paramagnetic regions. In two dimensions and at zero temperature,
the RBIM has a phase with spin glass correlations \cite{SG}.  The
three non-trivial fixed points along the full line are the pure Ising
fixed point, the Nishimori point {\it N} at the crossing with the
Nishimori line (dotted line) and the zero temperature point whose
properties are still mostly unknown.

For the $q$-state random Potts model with $q>2$, a Nishimori line can
also be obtained \cite{ns}. Reexpressing the disorder distribution as
\begin{equation}
P(J_{ij}) = p e^{K \delta^{(q)}(J_{ij})} \; \; \hbox{with} \; \;
K=\log \big( 1/p-(q-1) \big)
\label{newdist}
\end{equation}
the Nishimori line is given by the condition $K=\beta$ which
generalizes (\ref{NishimoriLine}) to $q > 2$.  We refer to
\cite{jacpic} for more details.

Note that the existence of a random fixed point is not related to the
existence of a Nishimori line. One could have chosen a distribution of
disorder for which no Nishimori line exist, still a random fixed point
would exist. This was observed for instance by S{\o}rensen {\em et
al.}\ \cite{Sorensen} for the $q=3$ random bond Potts model with a
distribution of disorder different from (\ref{olddist}). The main
advantage of the Nishimori line is that it allows to locate the random
fixed point while scanning only one parameter.

Over the last years, many numerical and analytical efforts have been
made in order to identify the universality class of the Nishimori
point. There is a long list of numerical results
\cite{MM,ON,SA,CF,OI,AQdS} with values for the critical exponents that
are very close to those of percolation

\section{Results}

We now turn to our numerical results. These were obtained by extensive
numerical transfer-matrix calculations of the Nishimori point with the
binary bond distribution (\ref{pmJprob}) for $q=2$ and with the
distribution (\ref{olddist}) for $q=3$ on the square lattice.

The first set of results is based on measurements of
the effective central charge, which is a very efficient way to locate
a fixed point. The effective central charge $c$ is obtained as the
universal coefficient of the finite-size correction to the free energy
{\it per site} for periodic boundary conditions \cite{central}
\begin{equation}
f_L^{(p)} = f_{\infty}^{(p)} + {c \pi \over 6 L^2} + \cdots.
\label{cDef}
\end{equation}
We first turn our attention to the determination of the properties of
the weak random point for $q=3$. The finite-size estimates for the
effective central charge {\em increase} along the massless RG flows
and approach the fixed point values for $L \to \infty$. Note that this
does not contradict Zamolodchikov's $c$-theorem \cite{Zamc} since the
present theories are not unitary. The Ferro-Para boundary
(cf.~Fig.~\ref{phaseRBI}b) can be traced by identifying the maximum of
the finite-size estimates of $c$ as a function of $T$, for various
fixed values of $p$.

Since the randomness is strong, and since the fits to (\ref{cDef})
must be based on at least two different sizes $L$ to eliminate the
non-universal quantity $f_{\infty}^{(p)}$, we have taken several
precautions \cite{jacpic} in order to obtain small error bars on
the $f_L^{(p)}$. Our strips have length $N=10^5$, and we average
$f_L^{(p)}$ over up to $10^5$ independent realizations.

The measurement of the effective central charge at the Ferro-Para boundary as
a function of $p$ shows the existence of an attractive fixed
point at $p \sim 0.04$ with a central charge slightly larger than
$c_{\rm pure} = 4/5$, characterizing the pure 3-state Potts
model. After extrapolation \cite{jacpic},
we arrive at the final result
\begin{equation}
 c_{F} = 0.8025(10) \, ,
\end{equation}
which compares favorably with the perturbative result $c_{\rm pert} =
0.8026 + {\cal O}\left((q-2)^5\right)$ \cite{ludwig2,Perturb} for the $q=3$
RBPM.

The next result is the numerical location of the Nishimori point along
the Nishimori line \cite{honpicpuj}. Again, this can be done by
measuring $c$ along the Nishimori line. For the $q=2$ case,
we have used the domain-wall free energy \cite{MM} to
locate the critical concentration of disorder $p_c$ since it provides
a better precision. For a strip of width $L$ the domain-wall
free energy $d_L$ is defined as (omitting a factor of $\beta$)
\begin{equation}
d_L = L^2 \left(f_L^{(p)} - f_L^{(a)}\right) \, ,
\label{defDW}
\end{equation}
where $f_L^{(p)}$ is the free energy per site of a strip of
width $L$ with periodic boundary conditions and $f_L^{(a)}$ the
corresponding one with {\it antiperiodic} boundary conditions.  $d_L$
is an observable which can be used directly to study the RG flow under
scale transformations. In particular, it is constant at a fixed point.

We have computed $f_L^{(p)} = {\ln Z^{(p)} \over L N}$ and $f_L^{(a)}
= {\ln Z^{(a)} \over L N}$ employing a standard transfer matrix
technique with sparse matrix factorization (see, e.g., \cite{Night})
on strips of length $N = 10^6$. Again, special tricks are used to
reduce fluctuations \cite{honpicpuj}.
We used around 1000 to 4000 samples of $L \times 10^6$ strips to
obtain sufficiently small error bars for $L \le 12$.

\begin{figure}[t]
\centerline{\psfig{file=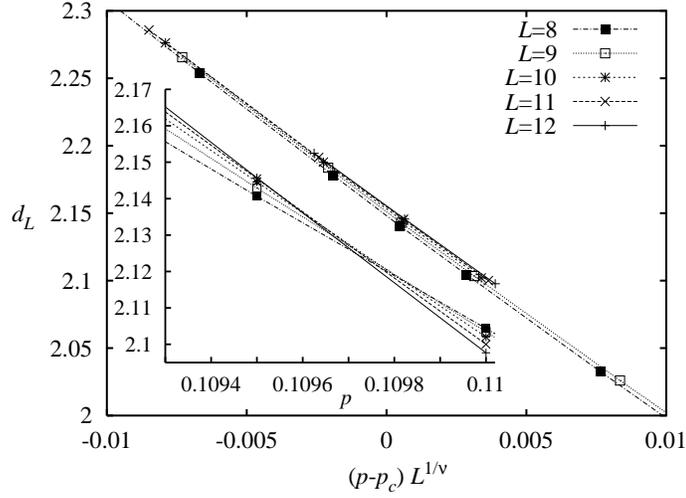,width=0.8\columnwidth,angle=270}}
\smallskip
\caption{Domain-wall free energy for $q=2$ along the
Nishimori line (\ref{NishimoriLine}). The inset shows the raw data and the
main panel the scaling collapse with $p_c = 0.1094$ and
$\nu = 1.33$.
\label{figScaleD}
}
\end{figure}

The inset of Fig.~\ref{figScaleD} shows $d_L(p)$ along the Nishimori
line (\ref{NishimoriLine}) in the vicinity of the critical
concentration $p_c$. A finite-size estimate for $p_c$ is given by the
crossing points $d_{L_1}(p_c) = d_{L_2}(p_c)$.  After extrapolation to
an infinitely wide strip, one obtains
\begin{equation}
p_c = 0.1094(2) \; .
\label{pcrit}
\end{equation}
This estimate improves upon the accuracy of earlier estimates
\cite{ON,SA,OI,AQdS}. It agrees perfectly with the transfer matrix
computations \cite{ON,AQdS}, in particular a very recent one \cite{MC},
while we find a slightly smaller value of $p_c$ than \cite{SA,OI}.

One an also extract the value of $\nu$ from (\ref{defDW}). We obtained
$\nu \approx 1.33$ \cite{honpicpuj}. Merz and Chalker \cite{MC} were able to
go to much larger strip widths and obtained the probably more accurate
result $\nu \approx 1.50$. While our estimate was compatible with
the value for percolation $\nu = 4/3$ \cite{StAh}, the one of \cite{MC}
is not.

Next, we discuss the effective central charge $c$ for $q=2$. One has
$c=1/2$ for the critical point of the pure Ising model, but it has not
been determined yet for the Nishimori point. In the process of
computing $d_L$ we have also obtained estimates of $f_L^{(p)}$ for
different values of $p$. One can either fit these values for
$f_L^{(p)}$ exactly by (\ref{cDef}) ignoring further corrections in
which case the data for the smallest values of $L$ should not be used.
Or one includes a correction term of the form $L^{-4}$ which improves
the convergence with system size. These two approaches yield
consistent estimates for a given $p$. In addition, one can test that
the result does not change significantly if other higher-order
corrections are added.  It should also be noted that the sensitivity
of the estimates for $c$ with respect to the location of $p_c$ is
negligible in comparison with the errors coming from the finite-size
analysis. The final result is that the following is a safe estimate
for $c$ at the Nishimori point of the $\pm J$ RBIM:
\begin{equation}
c = 0.464 (4) \, .
\label{cVal}
\end{equation}
This should be compared to the value for percolation over Ising
clusters $c= {5 \sqrt{3} \ln{2} \over 4 \pi} \approx 0.4777$
\cite{JC}. Even if our result (\ref{cVal}) is close to this value, it
appears safe to conclude that the Nishimori point is {\em not} in the
universality class of percolation, at least not the one expected from
Ising clusters.

The situation seems rather different for the case $q=3$. In that case,
the numerical location of the Nishimori point was only made by
measuring $c$ along the Nishimori line and determining the
point for which one obtains a maximum. We find
\cite{jacpic} a fixed point {\it N} at $p_{N} = 0.0785(10)$
with a central charge estimate
\begin{equation}
 c_{N} = 0.756(5) \, .
\end{equation}
This is in remarkable agreement with the value of the central charge
for the percolation limit in the $q=3$ RBPM: $c= \frac{5 \sqrt{3} \ln{3}}{4
\pi} \approx 0.7571$ \cite{JC}. Below we shall return to the question
whether the Nishimori point is ``just'' percolation. We should also
mention that for the case $q=3$ the error bars have been obtained by
extrapolating the data assuming that for each fit the relative
deviation from the infinite-size result is the same as in the pure
model. Taking a more conservative approach as we have done for the
case $q=2$ would produce larger error bars, but the conclusion would
remain the same.

Another important quantity is the magnetic exponent $\eta$. This
exponent can be measured, for example, by computing spin-spin
correlation functions. As mentioned earlier, along the
Nishimori line the moments of these correlation functions are equal
two by two (for $q=2$):
\begin{equation}
[\langle S(x_1,y_1) S(x_2,y_2)\rangle^{2k-1}]
= [\langle S(x_1,y_1) S(x_2, y_2)\rangle^{2k}]
\label{2by2}
\end{equation}
for any integer $k$. Here $[\cdots]$ stands for the average over the
disorder. Assume now that the correlation functions (\ref{2by2}) decay
algebraically on a plane and define by $x,y$ the coordinates on the
infinite cylinder of circumference $L$, with $x\in [1,L]$ and $y\in
]-\infty,+\infty[$.  Using a conformal mapping, one infers then the
following behavior of the correlation functions on the cylinder:
\begin{equation}
[\langle S(x_1,y) S(x_2,y)\rangle ^n] \propto
  \left(\sin\left({\pi (x_2-x_1)\over L}\right) L\right)^{- \eta_{n}} \, .
\label{fitsscf}
\end{equation}
For a pure system, one has $\eta_n = n \times \eta$. On the other
hand, in the case of percolation over Ising clusters,
the moments of spin correlation functions are all equal, whence
all $\eta_n = \eta$ at the critical point.

\begin{figure}[t]
\centerline{\psfig{file=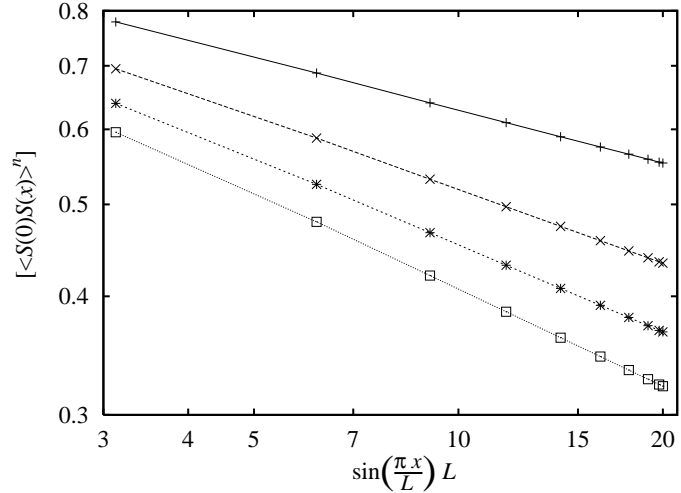,width=0.8\columnwidth,angle=270}}
\smallskip
\caption{Moments of the spin-spin correlation function for $q=2$,
$p=0.1095$ and $L=20$. We only show the odd moments: $n=1$ ($+$),
$n=3$ ($\times$), $n=5$ ($*$) and $n=7$ (open boxes). Error bars are
smaller than the size of the symbols. The values of the exponents are
given in (\ref{valExp}).
\label{ss}
}
\end{figure}

In order to verify this we have calculated the spin-spin correlation
functions on cylinders of width $L$ and length $400\times L$, ({\it
i.e.}\ with the length $\gg L$) for $L$ up to 20. We have checked that
for width $L=12$, lattice and finite length corrections are of order
1\%. One example of these correlation functions can be seen in Fig.~\ref{ss}
(for $q=2$ with $x_1=y=0$ and $x_2=x$) on a doubly
logarithmic scale. One observes that the correlation functions nicely
obey the power law (\ref{fitsscf}), thus verifying both the correct
location of the critical point as well as the functional form of the
spin-spin correlation function in a finite strip.

We can then fit the exponents by studying the dependence with distance
of the correlation functions (\ref{fitsscf}) or by studying the
dependence with $L$ for the fixed location $x = L/2$. The first method
has proven to give smaller error bars and we obtain for the family of
exponents $\eta_n$ for $p_c = 0.1095$ and $L=20$:
\begin{eqnarray}
\eta_{1}&=&\eta_{2}= 0.1854 ; \qquad \eta_{3}=\eta_{4}= 0.2561 \nonumber \\
\eta_{5}&=&\eta_{6}= 0.3015 ; \qquad \eta_{7}=\eta_{8}= 0.3354 \, ,
\label{valExp}
\end{eqnarray}
with relative errors at most of the order of 1\%.

One immediately notices two things:

{\it i)} The value for $\eta_1$ differs considerably from the value of
percolation $\eta = 5/24 \approx 0.2083$ (see, e.g., \cite{StAh}),

{\it ii)} the exponents for higher moments are also considerably
different from $\eta_1$ which is also clear from inspection of
Fig.~\ref{ss}.

We have also calculated estimates for the exponents assuming different
values for $p_c$, namely, $p_c = 0.109$ and $0.110$ and the results
are still distinct from the ones of percolation (we obtain $\eta_1 =
0.180(1)$ for $p=0.109$ and $\eta_1=0.190(1)$ for
$p=0.110$). Moreover, it is only in the region very close to
$p=0.1095$ that we obtain a stable estimate for $\eta_1$ as we
increase the width $L$ of the lattices. One can then conclude from the
exponents controlling the algebraic decay of the correlation functions
that the Nishimori point is not in the percolation universality class.

The same measurements were also performed in the case of $q=3$, for
which case the largest system size employed was $L=12$. Performing the
same fit of the data, we obtain the following exponents for $p=0.080$
\begin{eqnarray}
\eta_1 &=& 0.21239 (35) ; \qquad \eta_2=0.25192 (39) \nonumber \\
\eta_3 &=& 0.30824 (47) ; \qquad \eta_4= 0.33773 (52) \, ,
\end{eqnarray}
the corresponding values for $p=0.079$ being some 6\% smaller.

Although our value of $\eta_1$ is now consistent with percolation,
this scenario can be excluded by considering the higher moments.

\section{Conclusions}
In conclusion, we have studied the $\pm J$ random-bond Ising model as
well as a $q$-state (Potts-like) generalization that allows for the
definition of a Nishimori line. Both models possess a strong disorder
fixed point with multiscaling exponents different from those of
percolation, whereas a weak disorder fixed point that coincides with
that of the well-studied random-bond Potts model is also present in
the latter model.  In both cases, the effective central charge at {\it
N} is remarkably close to the percolation value $c= {5 \sqrt{3} \ln{q}
\over 4 \pi}$, which however appears to be ruled out numerically for
$q=2$. Open questions concerning our models include the study of their
zero-temperature limit, the possibility of reentrance, and of the
behavior for $q>4$.

M.~P. thanks the organizers of the NATO Advanced Research Workshop on
``Statistical Field Theories'', Como 18-23 June 2001 for the
invitation to participate and to present this work. We would like to
thank J.~Cardy, S.~Franz, J.~M.~Maillard, G.~Mussardo, N.~Read and
F.~Ritort for useful discussions and comments.


\begin{chapthebibliography}{1}

\bibitem{BPZ} A.~A.~Belavin, A.~M.~Polyakov and A.~B.~Zamolodchikov,
\newblock {\it Nucl. Phys.}~{\bf B241}, 333 (1984); \newblock {\it
J. Statist. Phys.}~{\bf 34}, 763 (1984).

\bibitem{Z} M. R. Zirnbauer, \newblock {\it J. Math. Phys.}~{\bf 37},
4986 (1996); A. Altland and M. R. Zirnbauer, \newblock {\it
Phys. Rev.}~B {\bf 55}, 1142 (1997).

\bibitem{JC} J. L. Jacobsen and J. Cardy, \newblock {\it
Nucl. Phys.}~{\bf B515}, 701 (1998).

\bibitem{MM} W. L. McMillan, \newblock {\it Phys. Rev.}~B {\bf 29},
4026 (1984).

\bibitem{DotDot} Vik.~S.~Dotsenko and Vl.~S.~Dotsenko, \newblock {\it
Sov. Phys. JETP Lett.}~{\bf 33}, 37 (1981); \newblock {\it
Adv. Phys.}~{\bf 32}, 129 (1983).

\bibitem{shalaev} B.~N.~Shalaev, \newblock {\it Sov. Phys. Solid
State}~{\bf 26}, 1811 (1984).

\bibitem{shankar} R.~Shankar, \newblock {\it Phys. Rev. Lett.}~{\bf
58}, 2466 (1987).

\bibitem{ludwig2} A.~W.~W.~Ludwig, \newblock {\it Nucl. Phys.}~{\bf
B285}, 97 (1987); \newblock {\it Nucl. Phys.}~{\bf B330}, 639 (1990).

\bibitem{ns} H.~Nishimori and M.~J.~Stephen, {\it Phys.~Rev.}~B {\bf
27}, 5644 (1983).

\bibitem{Gold} Y.~Y.~Goldschmidt, {\it J. Phys. A: Math. Gen.}\ {\bf
22}, L157 (1989).

\bibitem{Perturb} A.~W.~W.~Ludwig and J.~L.~Cardy, {\it
Nucl.~Phys.}~{\bf B285}, 687 (1987); Vl.~Dotsenko, M.~Picco and
P.~Pujol, {\it Nucl.~Phys.~}{\bf B455}, 701 (1995).

\bibitem{Sorensen} E.~S.~S\o rensen, M.~J.~P.~Gingras and D.~A.~Huse,
{\it Europhys.~Lett.}~{\bf 44}, 504 (1998).

\bibitem{N} H. Nishimori, \newblock {\it Prog. Theor. Phys.}~{\bf 66},
1169 (1981); \newblock {\it J. Phys. Soc. Jpn.}~{\bf 55}, 3305 (1986);
Y. Ozeki and H. Nishimori, \newblock {\it J. Phys. A: Math. Gen.}~{\bf
26}, 3399 (1993).

\bibitem{LG} A. Georges and P. Le Doussal, \newblock {\it unpublished
preprint} (1988); P.~Le Doussal and A.~B.~Harris, {\it
Phys. Rev. Lett.}~{\bf 61}, 625 (1988); {\it Phys. Rev.}~B~{\bf 40},
9249 (1989).

\bibitem{SG} I. Morgenstern and K. Binder, \newblock {\it
Phys. Rev.}~B~{\bf 22}, 288 (1980); I. Morgenstern and H. Horner,
\newblock {\it Phys. Rev.}~B~{\bf 25}, 504 (1982); W. L. McMillan,
\newblock {\it Phys. Rev.}~B {\bf 28}, 5216 (1983).

\bibitem{jacpic} J.~L.~Jacobsen and M.~Picco, {\it Phys.~Rev.}~E, to
be published [cond-mat/0105587].

\bibitem{ON} Y.\ Ozeki and H.\ Nishimori, \newblock {\it J.\ Phys.\
Soc.\ Jpn.}\ {\bf 56}, 3265 (1987).

\bibitem{SA} R. R. P. Singh and J. Adler, \newblock {\it Phys. Rev.}~B
{\bf 54}, 364 (1996).

\bibitem{CF} S. Cho and M. P. A. Fisher, \newblock {\it Phys. Rev.}~B
{\bf 55}, 1025 (1997).

\bibitem{OI} Y.\ Ozeki and N.\ Ito, \newblock {\it J.\ Phys.\ A:
Math.\ Gen.}\ {\bf 31}, 5451 (1998).

\bibitem{AQdS} F. D. A. Aar\~ao Reis, S. L. A. de Queiroz and R. R. dos
Santos, \newblock {\it Phys. Rev.}~B {\bf 60}, 6740 (1999).

\bibitem{central} H. W. J. Bl\"ote, J. L. Cardy and M. P. Nightingale,
\newblock {\it Phys.\ Rev.\ Lett.}\ {\bf 56}, 742 (1986); I.\ Affleck,
\newblock {\it Phys.\ Rev.\ Lett.}\ {\bf 56}, 746 (1986).

\bibitem{Zamc} A.~B.~Zamolodchikov, {\it Pis'ma
Zh.~Eksp.~Teor.~Fiz.}~{\bf 43}, 565 (1986) [{\it JETP Lett.}~{\bf 43},
730 (1986)].

\bibitem{honpicpuj} A.\ Honecker, M.\ Picco and P.\ Pujol, {\it
Phys.~Rev.~Lett.}~{\bf 87}, 047201 (2001).

\bibitem{Night} M. P. Nightingale, pp. 287-351 in: V. Privman (ed.),
{\it Finite Size Scaling and Numerical Simulations of Statistical
Physics}, World Scientific, Singapore (1990).

\bibitem{MC} F. Merz and J. T. Chalker, cond-mat/0106023.

\bibitem{StAh} D.\ Stauffer and A.\ Aharony, {\it Introduction to
Percolation Theory}, 2nd edition, Taylor \& Francis, London (1994).

\end{chapthebibliography}
\end{document}